\documentclass[11pt,aps,prd,nofootinbib,superscriptaddress, onecolumn,preprintnumbers,balancelastpage]{revtex4-1}
\usepackage{amssymb,amsmath,latexsym,graphics, graphicx,epsfig,multirow,comment,hyperref,appendix, verbatim} 

\def\OO{{\cal O}}

\newcommand{\vmin}{v_{\rm min}}
\newcommand{\Nperi}{N_{\rm peri}}

\newcommand{\gsim}{\lower.7ex\hbox{$\;\stackrel{\textstyle>}{\sim}\;$}}
\newcommand{\lsim}{\lower.7ex\hbox{$\;\stackrel{\textstyle<}{\sim}\;$}}
\newcommand{\kms}{\text{ km s${}^{-1}$}}
\newcommand{\Msun}{\text{ M}_{\odot}}

\usepackage[usenames,dvipsnames]{color}
\definecolor{orange}{RGB}{220,120,0}

\begin{document}

\title{Direct Detection of Dark Matter Debris Flows}

\author{Michael Kuhlen}
\affiliation{Theoretical Astrophysics Center, University of California, Berkeley, CA 94720}

\author{Mariangela Lisanti}
\affiliation{PCTS, Princeton University, Princeton, NJ 08544}

\author{David N. Spergel}
\affiliation{Department of Astrophysical Sciences, Princeton University, Princeton, NJ 08540}

\begin{abstract}
Tidal stripping of dark matter from subhalos falling into the Milky
Way produces narrow, cold tidal streams as well as more spatially
extended ``debris flows'' in the form of shells, sheets, and plumes.
Here we focus on the debris flow in the Via Lactea II simulation, and
show that this incompletely phase-mixed material exhibits distinctive
high velocity behavior. Unlike tidal streams, which may not
necessarily intersect the Earth's location, debris flow is
spatially uniform at 8 kpc and thus guaranteed to be present in the
dark matter flux incident on direct detection experiments. At
Earth-frame speeds greater than \mbox{450 km/s}, debris flow
comprises more than half of the dark matter at the Sun's location, and
up to 80\% at even higher speeds. Therefore, debris flow is
 most important for experiments that are particularly sensitive to the
high speed tail of the dark matter distribution, such as searches
for light or inelastic dark matter or experiments with directional
sensitivity. We show that debris flow
 yields a distinctive recoil energy spectrum and a broadening of the
distribution of incidence direction.
\end{abstract}


\maketitle

\section{Introduction}

The Galactic dark matter halo forms through a process of hierarchical structure formation, with smaller halos merging together to form a larger host.  This process of halo accretion occurs over the span of billions of years.  Dark matter that merged early on virializes and is smoothly distributed at present.  However, more recent mergers can leave relic structures in the Milky Way, observed as features in the spatial and velocity distribution of Galactic halo stars.  Understanding the origin of these features is critical for piecing together the formation history of our Galaxy, and can provide clues for distinctive signatures to search for with dark matter experiments. 
  
The dark matter in the solar neighborhood is commonly assumed to be smoothly distributed in space and to have a Maxwellian velocity distribution~\cite{Drukier:1986tm, Lewin:1995rx}.  High resolution numerical simulations of the hierarchical formation of Milky-Way-like dark matter halos, however, predict large of phase-space substructure throughout the halo \cite{Diemand:2008in,Zemp:2008gw,Springel:2008cc,Vogelsberger:2008qb}.  This is in agreement with collisionless dynamics and Liouville's theorem, which imply that the initial cold dark matter three-dimensional phase-space manifold\footnote{In Cold Dark Matter theory, the thermal velocity dispersions are close to zero, yielding a very thin three-dimensional sheet in phase space as an initial configuration.}  evolves in a continuous manner by folding and stretching, but never tearing. Gravitationally bound subhalos, such as those thought to host the Milky Way dwarf satellite galaxies, are examples of \textit{spatial} phase-space substructure. The current census of Milky Way dwarf satellite galaxies stands at 22, but many more are likely to be discovered with future surveys \cite{Tollerud:2008ze}.  Additionally, the simulations predict the presence of many thousands of dark subclumps within the Milky Way's virial volume, too small to host a luminous stellar component, but potentially interesting dark matter annihilation sources \cite[e.g.,][]{Kuhlen:2008aw,Kuhlen:2009kx}. The substructure abundance relative to the smooth host halo mass distribution is found to decrease towards the Galactic Center, a natural consequence of the stronger disruptive tidal forces and shorter dynamical times closer to the center of the potential. Current estimates based on the results from the highest resolution numerical simulations find that spatial substructure is unlikely to significantly modify the local dark matter density at 8 kpc \cite{Kamionkowski:2008vw,Vogelsberger:2008qb,Kamionkowski:2010mi}. Barring drastic changes in the properties of very low-mass subhalos,\footnote{Calculations of the contribution of local substructure to $\bar{\rho}$(8 kpc) must extrapolate subhalo scaling relations for many orders of magnitude below the simulations' resolution limit.} the assumption of a smooth dark matter distribution at 8 kpc appears to be a good one.

The situation is quite different for velocity substructure. The same tidal disruption processes that render the local dark matter distribution spatially smooth are sources of velocity substructure. Indeed, the speed distributions measured in high resolution numerical simulations exhibit deviations from the standard Maxwellian assumption, especially at large speeds~\cite{Kuhlen:2009vh, Vogelsberger:2008qb, Lisanti:2010qx}. As we show below, the vast majority of high-speed dark matter particles in the solar neighborhood have been recently accreted and are partially phase-mixed, having not yet come into equilibrium with the rest of the halo. We can further distinguish between velocity structure that is spatially localized, such as tidal streams, and that which is spatially homogenized, which we designate as ``debris flow'' \cite{Lisanti:2011as}. Both streams and debris flow arise from the disruption of satellites that fall into the Milky Way, but differ in the relative amount of phase-mixing that they have undergone. 

Tidal streams consist of material that has been stripped from an infalling satellite, and that has not yet had the time to spatially mix. It is dynamically cold (meaning its internal velocity dispersion is much less than the Milky Way halo's), and it is still spatially confined to a narrow stream with a one-dimensional morphology. There are several known examples of stellar tidal streams in the Milky Way halo. One of the most dramatic examples is the Sagittarius stream~\cite{Ivezic:2000ua, Yanny:2000ty}, which is clearly associated with the on-going tidal disruption of the Sagittarius dwarf galaxy~\cite{Johnston:1995vd}.  The SDSS ``Field of Streams'' \cite{Belokurov:2006ms, Belokurov:2006kc} contains a number of additional tidal stream candidates such as the Monoceros and Orphan streams. The existence of stellar streams associated with disrupting dwarf galaxies implies that dark matter streams should have formed by a similar mechanism.  The presence of such a stream in the local neighborhood could significantly affect the predictions of event rates and recoil spectra at direct detection experiments \cite{Alves:2010pt, Lang:2010cd, Gondolo:2005hh, Kuhlen:2009vh}. While dark matter streams have been identified in N-body simulations \cite{Zemp:2008gw,Vogelsberger:2008qb,Kuhlen:2009vh,Maciejewski:2010gz,Elahi:2011dy}, the probability that a single stream dominates the local dark matter density is less than 1\%~\cite{Vogelsberger:2008qb}.

In this paper, we instead focus on debris flow, which represents a more ubiquitous type of velocity substructure. The term ``debris flow" refers to the sum total of all material stripped from infalling subhalos that has not completely phase-mixed. As such, it comprises dynamically cold and narrow tidal streams from recently infalling subhalos, older tidal streams that have been wrapped a number of times, as well as material that was lost from halos in the form of sheets and plumes in the violent gravitational shocks experienced at pericenter passages \cite{Choi:2008xi}. Rather than considering multiple debris flows, each associated with an individually disrupting subhalo, we instead view debris flow as a single feature of the velocity distribution.

The salient difference between an individual tidal stream and debris flow is that the former is dynamically cold and has a one-dimensional morphology, while the latter is dynamically hot and is spatially ubiquitous in the central regions of the Milky Way halo. For a collisionless system, any diffusion in configuration space must be accompanied by a decrease in the width of the velocity distribution in order to preserve phase-space density.  Therefore, one might expect debris flow to be \textit{colder}, not hotter, than tidal streams.  However, because debris flow is the superposition of tidal debris from many disrupting satellites, its velocity dispersion is due to the relative velocity of material stripped from distinct subhalos as well as from the intersections of a single halo's tidal stream with itself.  While the individual fine-grained components of the tidal debris must be very cold, in aggregate they appear dynamically hot. As such, debris flow is velocity structure that is intermediate between the fully equilibrated host halo and dynamically cold and narrow tidal streams. 

In~\cite{Lisanti:2011as}, we used the Via Lactea-II simulation~\cite{Diemand:2008in, Kuhlen:2008qj} to study one sub-component of the debris flow, namely the portion that was bound to halos at the time of reionization ($z \sim 9$). Already this component exhibited an interesting speed distribution strongly peaked at $\sim 340$ km/s in the Galactic frame, quite unlike that of the underlying relaxed host halo. Here we extend this analysis by following a larger sample of subhalos throughout their entire accretion history. This allows us to get a better understanding of the origin and make-up of the debris flow.

Owing to its spatial homogeneity, this debris flow is guaranteed to be present in the solar neighborhood, and it is therefore very important to understand its implications for direct detection experiments, which are sensitive to the local distribution of dark matter velocities~\cite{Lewin:1995rx}.  These experiments consist of shielded detectors that measure the recoil energies of target nuclei scattering off dark matter particles passing through the Earth~\cite{Gaitskell:2004gd}.  The expected recoil spectrum is different for dark matter that is in velocity substructure rather than in the equilibrated component of the halo.  We will show in \S~\ref{sec: model} that debris flow results in a distinctive recoil spectrum, with more high energy events than is typically expected from the canonical Maxwellian velocity distribution.  These differences may be important in ameliorating the current tension between experiments, some of which have been observing anomalous signals~\cite{Bernabei:2008yi, Bernabei:2010mq, Aalseth:2010vx, Aalseth:2011wp, Angloher:2011uu} while others have not~\cite{Ahmed:2012vq, Aprile:2011hi, Ahmed:2009zw, Akerib:2005kh, Edelweiss:2011cy, Angle:2007uj, Angle:2009xb,Angloher:2004tr,Alner:2007ja, Lebedenko:2008gb,Akimov:2011tj,Ahmed:2010wy, Collar:2011kf,Angle:2011th}.

This paper is organized as follows. In \S~\ref{sec: debris_identification}, we describe in detail our procedure for identifying debris flow particles in the Via Lactea II simulation, and show that debris flow dominates the local dark matter distribution at high speeds. In \S~\ref{sec: debris_formation}, we present the speed distribution of the debris flow and discuss its origin. In \S~\ref{sec: model}, we go on to explore the implications of debris flow for direct detection experiments. This section also includes a simple model that accurately captures the phenomenology, and which can be used to model debris flow effects without resorting to high resolution numerical simulations. Finally, in \S~\ref{sec: conclusion},we present a brief discussion of the results and a conclusion.

\section{Identification of Debris Flow Particles}
\label{sec: debris_identification}

We use the Via Lactea-II (VL2) N-body simulation \cite{Diemand:2008in,Kuhlen:2008qj} to study the formation of debris flow in the Milky Way. VL2 is one of the highest resolution cosmological dark-matter-only simulations of the formation of a galactic halo. It resolves the virial volume of a Milky-Way-sized halo with about 1 billion particles of mass $4.1\times10^3 \Msun$ embedded in a cubic volume of 40 Mpc per side. The simulation is initialized at redshift 104.3 assuming a WMAP3 $\Lambda$CDM cosmology~\cite{Spergel:2006hy}, and evolved to the present. VL2 resolves large amounts of substructure in the Galactic halo, including subhalos and dark matter streams \cite{Zemp:2008gw,Kuhlen:2009vh}. Throughout the evolution, 400 full outputs, equally spaced in time, were written to disk.  

The 6DFOF halo-finding algorithm \cite{Diemand:2006ey} was used to identify the tightly bound centers of all \mbox{(sub-)halos} at 27 of the outputs (roughly every $\sim 680$ Myr). These centers were used to construct spherical density profiles, from which we calculated subhalo properties like $V_{\rm max}$, $R_{\rm Vmax}$, and the tidal radius (or $R_{200}$ for isolated halos) and corresponding mass. The subhalos are linked through time in two sets of evolutionary tracks. The first one ($T_0$) starts with the 20\,048 subhalos that have an identifiable remnant at $z=0$ and reached $V_{\rm max} > 4$ km/s at some time, and traces their most massive progenitor halo \textit{backwards through time} from $z=0$. The second ($T_{4.56}$) starts with the 20\,000 most massive halos at $z=4.56$ and traces their descendant halos \textit{forward through time}. The overlap between the two tracks consists of 11\,870 subhalos, and 7\,433 subhalos in $T_{4.56}$ do not have a $z=0$ remnant. Both sets of tracks (as well as additional data) are available at the Via Lactea Project webpage \cite{VLwebsite}. For the 20\,000 subhalos in $T_{4.56}$, we additionally traced the 6DFOF-linked central particles through the intermediate outputs, so we have orbital information (positions and velocities) at all 400 outputs (every $\sim$ 34 Myr).

A dark matter particle is labeled as ``debris'' if it was bound to some halo at $z>0$ and is no longer bound to any halo but the host today. Operationally, we restrict our analysis to the 4\,232\,452 particles located between 7.5 and 9.5 kpc at $z=0$, and determine for each of these particles to which (sub-)halos it is bound at every one of the 27 coarse outputs. For every particle, we determine whether it is debris, and, if so, at what redshift it was stripped off its birth halo.  Because our halo finding procedure does not ``un-bind'' particles, we consider a particle to be bound to a halo if it lies within the halo's tidal radius. This may slightly overestimate the amount of debris, because a small fraction of particles are assigned halo membership even though they are just passing through, but we have explicitly checked that this is not a large effect for a small subsample of halos, and do not expect this slight overestimate to affect our conclusions.

\begin{figure}[tb] 
   \centering
   \includegraphics[width=3.5in]{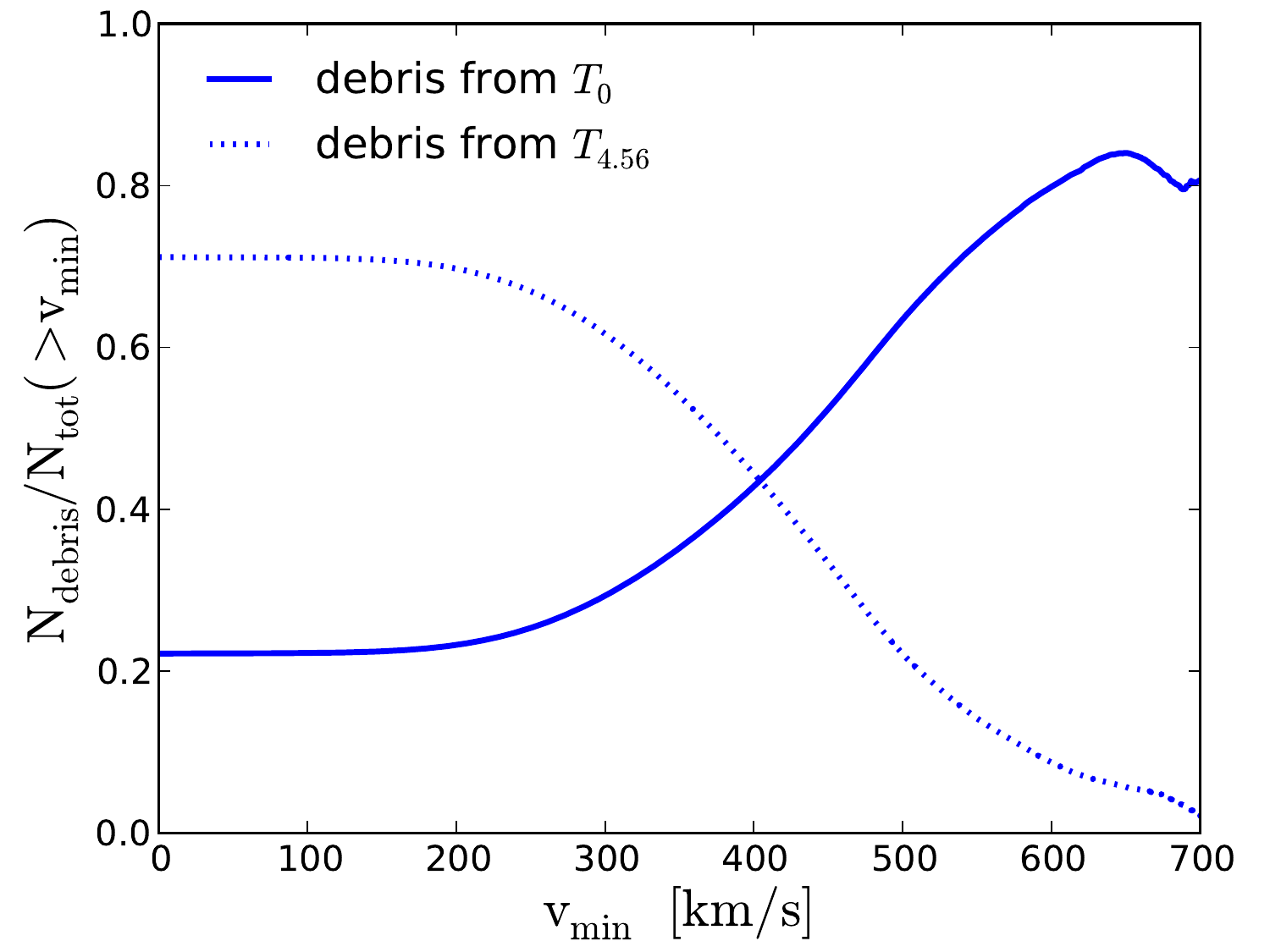} 
   \caption{Fractional density of debris particles above some minimum speed, $v_{\text{min}}$, in the Earth's rest frame (in June). The solid line is for debris particles with a $z=0$ remnant halo (from tracks $T_0$), and the dotted line for high redshift debris from halos that are completely disrupted by $z=0$. }
   \label{fig: gvminplot}
\end{figure}  

We construct debris catalogs from both $T_0$ and the subset of subhalos in $T_{4.56}$ that do not have a $z=0$ remnant. Note that this constitutes a marked improvement over our earlier study \cite{Lisanti:2011as}, in which we considered only particles that were bound around the time of reionization $z \sim 9$. Figure~\ref{fig: gvminplot} shows the fractional contribution ($N_{\rm debris} / N_{\rm tot}$) above a given Earth rest-frame speed. In total ($\vmin=0$ km/s), about 90\% of all particles at 8 kpc are debris, with 70\% having been stripped from halos that were completely disrupted prior to $z=0$, and 20\% from halos with surviving remnants. At higher $\vmin$, the relative contributions are reversed: debris from surviving subhalos exceeds debris from fully disrupted subhalos at 400 km/s, and makes up more than half of all the material at $\vmin > 450$ km/s.  Debris from surviving subhalos contributes as much as 85\% of the local material at $\vmin=650$ km/s. The $T_0$ debris curve is well fit by a Gauss error function,
\begin{equation}
\epsilon(E_R) \, \simeq \, 0.22 + 0.34 \, \left[ {\rm erf}\!\left(\frac{v_{\rm min} - 465 \, {\rm km/s}}{185 \, {\rm km/s}} \right) + 1 \right].
\end{equation}

\section{Formation Process of Debris Flows}
\label{sec: debris_formation}

\begin{figure}[htp] 
   \centering
\includegraphics[width=6.5in]{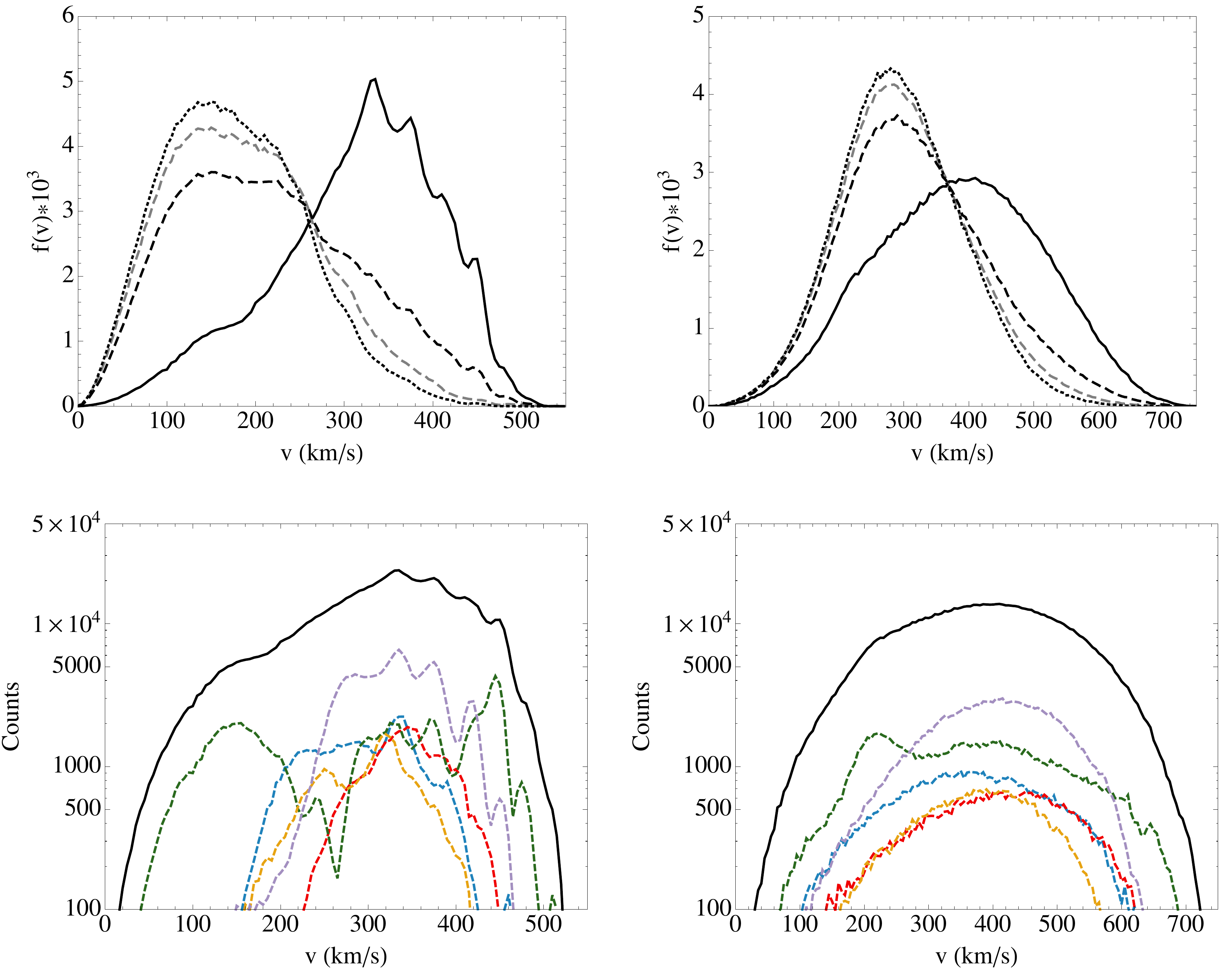}
   \caption{Top : Normalized speed distributions for debris from subhalos that are still present at $z=0$ (solid line), from subhalos present at $z=4.56$ but not at $z=0$ (dotted line), for all particles in the Milky Way (black dashed line), and for non-debris particles (gray dashed line). The comparison is made for particles in the radial shell $7.5 < r < 9.5$ kpc.  Bottom: Histogram of speed distribution for the debris flow (solid black), as well as the distributions of particles from a sample of subhalos that contribute the most to the debris flow (colored dashed: 19765:purple, 19624:green, 17928:blue, 17689:red, 18506:yellow).  The left panel shows the distributions in the Galactic frame, while the right panel is in the Earth frame (assuming $t_{\text{max}}$ = June 2).}
   \label{fig: fvdistribution}
\end{figure}

The left panel of Figure~\ref{fig: fvdistribution} shows the Galactic rest-frame speed distribution of the debris particles in the radial shell $7.5 < r < 9.5$ kpc, compared to the distribution of all particles, as well as non-debris particles, in the same radial shell. Note that these distributions are \textit{separately normalized} in order to highlight the difference in their shapes, but as a result their heights do not reflect the relative contributions of each component (see Figure~\ref{fig: gvminplot} for that information). The total distribution (black dashed) exhibits the well-known \cite{Diemand:2004kx,Hansen:2005yj,Vogelsberger:2008qb,Kuhlen:2009vh, Fairbairn:2008gz, Catena:2011kv} departures from the shape of a Maxwellian distribution, consisting of a deficit near the peak and an excess at high speeds.
The speed distribution for non-debris particles (grey dashed) is similar to the distribution for debris from fully disrupted subhalos (dotted), indicating that the $T_{4.56}$ debris has equilibrated with the host halo.  In contrast, the debris from surviving subhalos has an intriguing high-speed behavior, with a distribution (solid) peaked at $\sim 350$ km/s.  This is consistent with the results of~\cite{Lisanti:2011as}, which considered only a subset of particles contributing to the debris flow. For the remainder of this paper we consider the debris from fully disrupted subhalos to be part of the background halo, and henceforth the term ``debris flow'' will refer to debris from subhalos with a surviving $z=0$ remnant only.

On the right side of Figure~\ref{fig: fvdistribution}, we show the corresponding distributions shifted into the Earth's frame. These distributions are obtained by applying a Galilean boost of 
\begin{equation}
\vec{v}_e(t) = \vec{v}_{\text{LSR}} + \vec{v}_{\text{pec}} +  \vec{v}_{\oplus}(t),
\label{eq: vEarth}
\end{equation}
where $\vec{v}_{\text{LSR}}=(0, 220, 0) \text{ km/s}$ is the velocity of the local standard of rest (LSR)~\cite{Majewski:2008pz}, $\vec{v}_{\text{pec}} = (10, 5.23, 7.17) \text{ km/s}$ is the Sun's peculiar velocity with respect to the LSR~\cite{Dehnen:1997cq}, and $\vec{v}_{\oplus}$ is the velocity of the Earth in the Sun's rest frame, as specified in~\cite{Savage:2008er, Gelmini:2000dm}.  These velocities are taken in the coordinate system where $\hat{x}$ points towards the Galactic center, $\hat{y}$ points in the direction of Galactic rotation, and $\hat{z}$ points towards the Galactic north pole.  
These coordinates are associated with the ($v_r, v_{\theta}, v_{\phi}$) coordinates of the VL2 particles, for an arbitrary assignment of the Galactic plane.  The right side of Fig.~\ref{fig: fvdistribution} shows that the transformation into the Earth frame smooths out some of the peaks in the debris flow distribution observed in the Galactic rest-frame.\footnote{This smoothing arises because, in the transformation from Galactic to Earth frame, particles with different Galactic frame speed can end up with the same Earth frame speed, depending on their direction with respect to the Earth.} The debris flow distribution, however, maintains a significantly different shape and is shifted towards higher speeds.

\begin{table}
  \begin{tabular}{  | c | c | c | c | c | c | c | c | }
    \hline
    Subhalo ID & Mass ($z=0$) & $R_{\rm gc} (z=0)$ & Infall Mass & $z_{\rm infall}$ & $N_{\rm peri}$ & min($D_{\rm peri}$) & $f_{\rm debris}$ \\
               &  [$\Msun$]    & [${\rm kpc}$]    & [$\Msun$]   &             &             & [${\rm prop. kpc}$]    &                 \\
    \hline
    \hline              
    19765 & $9.8 \times 10^6$ & 20.9 & $4.1 \times 10^9$ & 1.9 & 12 & 4.1 & $1.2 \times 10^{-1}$ \\
    19624 & $5.8 \times 10^8$ & 21.8 & $2.7 \times 10^{10}$ & 1.6 & 6 & 6.6 & $9.3 \times 10^{-2}$ \\
    17928 & $5.7 \times 10^7$ & 42.3 & $5.8 \times 10^9$ & 2.9 & 15 & 5.9 & $4.5 \times 10^{-2}$ \\
    17689 & $1.2 \times 10^7$ & 44.6 & $7.9 \times 10^9$ & 2.9 & 15 & 3.7 & $3.2 \times 10^{-2}$ \\
    18506 & $4.3 \times 10^6$ & 34.1 & $1.1 \times 10^9$ & 3.6 & 21 & 2.4 & $2.8 \times 10^{-2}$ \\
    \hline
    18646 & $2.9 \times 10^8$ & 41.0 & $2.5 \times 10^9$ & 1.3 & 4 & 44 & $1.3 \times 10^{-3}$ \\
    \hline
  \end{tabular}
  \caption{The top five subhalos contributing the most mass to the debris flow, plus one fairly massive halo that contributes only very little.  The table lists each subhalo's ID, its $z=0$ mass and Galacto-centric radius ($R_{\rm gc}$), its mass at infall (first crossing of the host's $R_{\rm vir}$) and the redshift this occurred ($z_{\rm infall}$), the number of pericenter passages ($N_{\rm peri}$) and minimum pericenter distance ($D_{\rm peri}$) of its orbit, as well as the mass fraction of debris ($f_{\rm debris}$) it contributes.}
  \label{table: stats}
\end{table}
\begin{figure}[tp] 
   \centering
   \includegraphics[height=0.70\textheight]{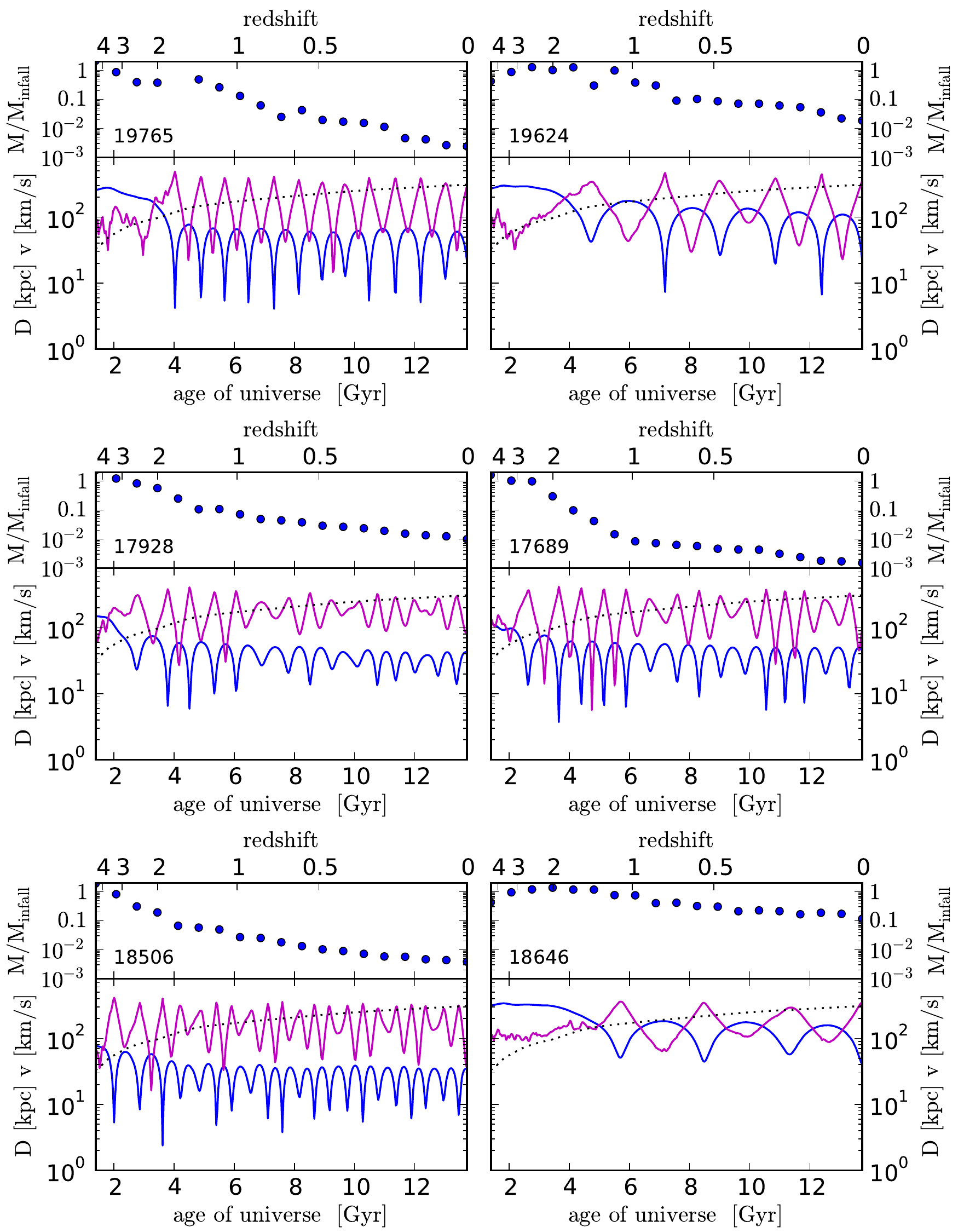} 
   \caption{Mass ratio ($M/M_{\rm infall}$; blue dots), Galacto-centric distance (blue lines) and relative speed (magenta lines) as a function of time for the five subhalos contributing the most mass to the debris flow, and one additional halo (bottom right panel) contributing only very little, $f_{\rm debris} = 9.8 \times 10^{-4}$. See Table~\ref{table: stats} for more information about these subhalos. Subhalo masses have only been determined at coarsely spaced outputs (every $\sim 680$ Myr), but the orbits of the subhalos' most strongly bound central particles (i.e. 6DFOF-linked) have been traced in the intermediate outputs (every $\sim 34$ Myr). The dotted line indicates the virial radius of the host halo.}
   \label{fig: orbits}
\end{figure}

The presence of several discrete sub-peaks in the debris distribution (e.g., at 330, 380, 420, and 460 km/s in the Galactic frame distribution) hints at the importance of a few individual halos. This is confirmed in the lower left panel of Figure~\ref{fig: fvdistribution}, in which we show the debris flow distribution on a logarithmic scale and over-plot the corresponding distributions (normalized to total debris flow) for the five subhalos contributing the most mass to the debris flow. Although no one subhalo dominates the peak at $\sim 350$ km/s, individual features are easily identified as being associated with one of these halos: the broad shoulder at $\sim 100$ km/s as well as the peak at 460 km/s, for example, are contributed by halo 19624, and the peaks at 330, 380, and 420 km/s come from halo 19765. In total, these five subhalos make up 31.8\% of the debris flow. Figure~\ref{fig: orbits} shows their mass loss and orbital information (Galacto-centric distance and speed) as a function of time, and their properties are summarized in Table~\ref{table: stats}. The number of pericenter passages undergone by these subhalos is strikingly high. With the exception of 19624 ($\Nperi=6$), they all have experienced more than 10 pericenter passages, and subhalo 18506 had more than 20. For comparison, the mean number of pericenter passages for all $T_{4.56}$ subhalos with at least one pericenter passage is only 4.3. The top five debris-contributing subhalos all have multiple \textit{deep} pericenter passages, reaching considerably below 10 kpc, which enables them to contribute to the local debris flow at 8 kpc.

The high number and depth of their pericenter passages is reflected in a large amount of mass loss. From first infall until $z=0$, these five subhalos have lost between 97.9\% and 99.7\% of their infall mass, and it is this material that makes up their contribution to the debris flow. Note that their mass loss is strongest in the earlier parts of their orbits \cite{Kravtsov:2004cm,Diemand:2007qr}.  As a contrast, we show in the bottom right panel of Figure~\ref{fig: fvdistribution} the orbital information for a sixth halo, which is representative of the population of subhalos that only contributes weakly to the debris. This subhalo has undergone a smaller number of pericenter passages, none of which reach closer than 44 kpc from the center, and it has lost less than 90\% of its mass. 

The speed of $\sim 350$ km/s at which the main debris flow peak occurs is easily explained by energy conservation, as it simply reflects the speed of the debris particles' parent subhalos orbiting in the Galactic potential. The five representative subhalos discussed above have infall redshifts between 3.6 and 1.6, and initial apocenter distances ranging from 74 to 176 kpc, with a mean of $\langle D_{\rm apo}^i \rangle = 96$ kpc. As the subhalos orbit in the host halo, they lose mass from tidal stripping and their orbits shrink due to the influence of dynamical friction and in response to the steadily growing mass of the host halo interior to their orbits. This shrinking continues until they become so light that dynamical friction ceases to be efficient ($M_{\rm sub} / M_{\rm host} \lesssim 10^{-2}$, \cite{BoylanKolchin:2007ku}) and the host halo's mass accretion halts \cite{Diemand:2007qr}. The apocenters of our five subhalos shrink by a factor of $D_{\rm apo}^f / D_{\rm apo}^i = 0.47$ to 0.78, with a mean final apocenter of $\langle D_{\rm apo}^f \rangle = 59$ kpc. At this distance, they have a mean speed of $\langle v_{\rm apo} \rangle = 54$ km/s. The difference in the late time ($z<1$) host halo potential between 59 kpc and 8.5 kpc is $6.7 \times 10^4$ (km/s)$^2$, and hence conservation of energy implies a mean speed at 8.5 kpc of 370 km/s, which is in very good agreement with the peak speed of the debris flow. These results also correspond well with the energy-infall relation recently elaborated on by~\cite{Rocha:2011aa}.

\begin{figure}[b] 
   \centering
   \includegraphics[height=2.7in]{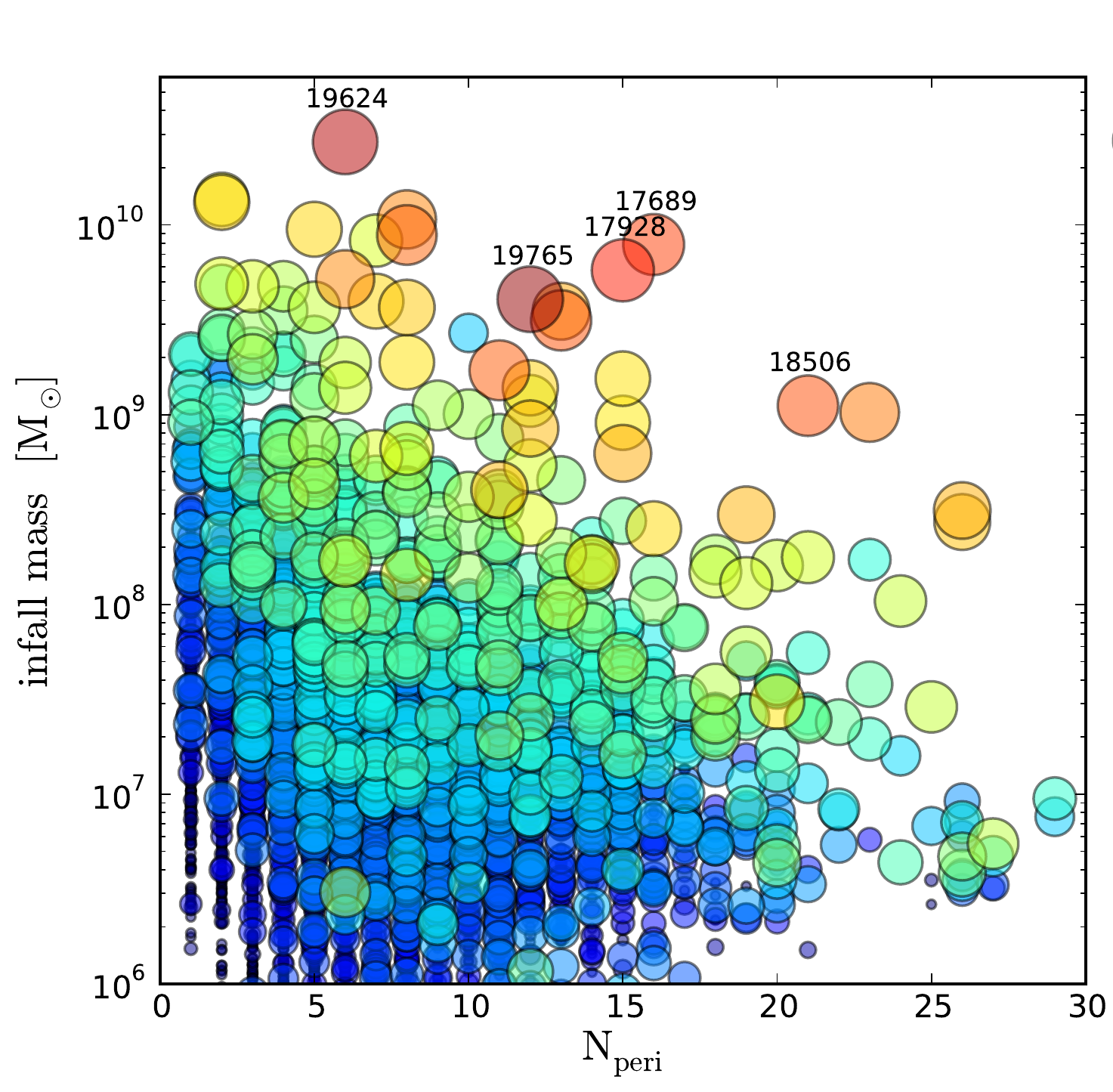}
   \includegraphics[height=2.7in]{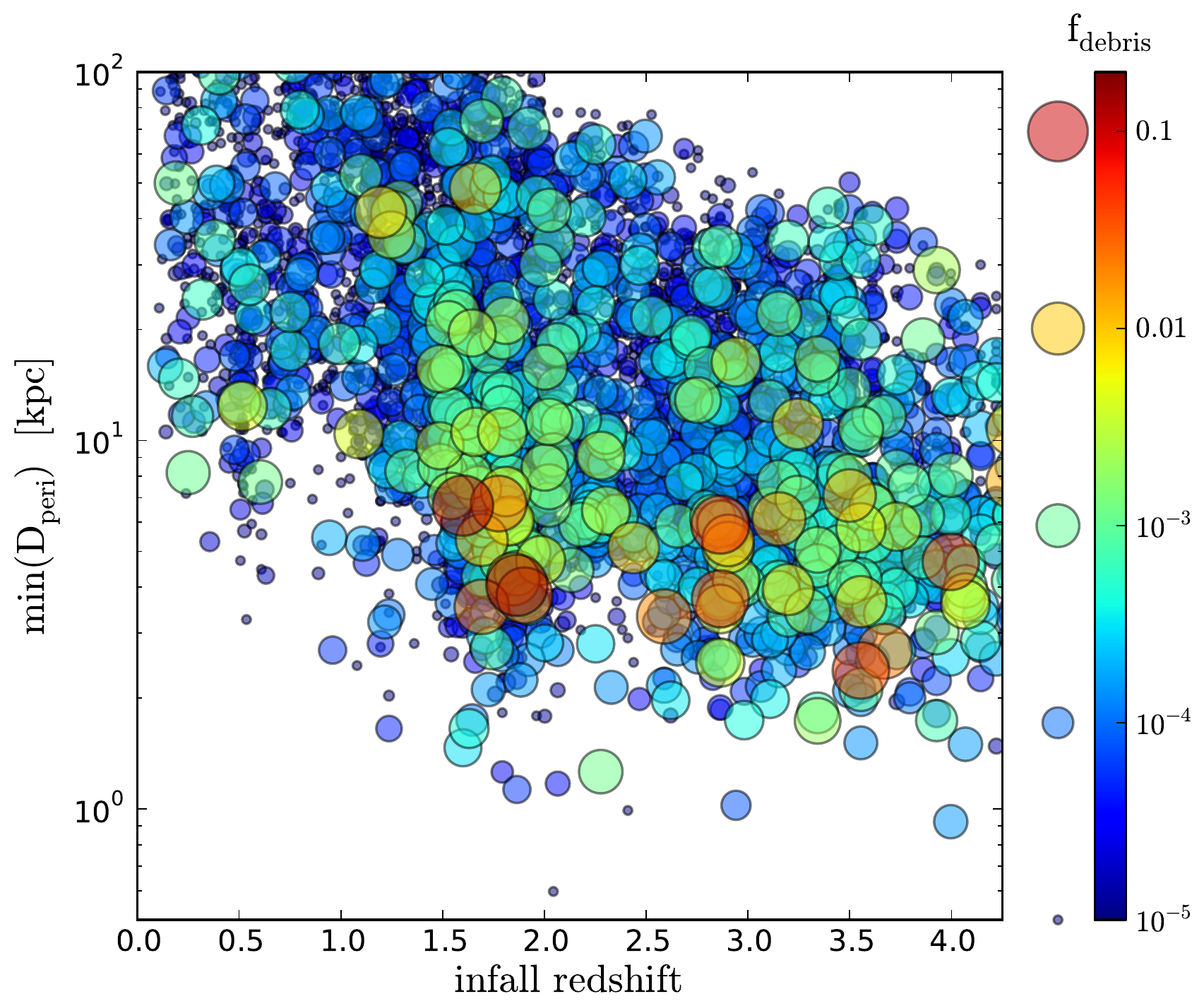}
   \caption{Scatter plots of infall mass vs. number of pericenter
     passages (\textit{left}) and closest pericenter approach
     vs. infall redshift (\textit{right}). The size and color of the
     symbols indicate the fractional contribution that a subhalo makes
     to the debris flow. Only subhalos with surviving remnants at
     $z=0$ are plotted.}
   \label{fig: debris_population}
\end{figure}

In Figure~\ref{fig: debris_population}, we extend our analysis to the full set of subhalos contributing to the debris flow. In the left panel, we show a scatter plot of the subhalos' infall mass versus their number of pericenter passages ($\Nperi$). The size and color of the symbols represent the fraction of the debris flow that a given subhalo contributes. In the right panel, we plot the distance of the deepest pericenter approach (min($D_{\rm peri}$)) against the subhalo's infall redshift. These two plots clearly demonstrate that the trends observed for the top five contributing subhalos continue to hold for the entire population. The amount of material contributed to the debris flow tends to increase with increasing infall mass, with larger $\Nperi$, and with decreasing min($D_{\rm peri}$) of the subhalo. The largest fraction of debris flow is contributed by subhalos accreting between $z=1.5$ and 4, and about 40\% of the debris is contributed by halos brought in with the last major merging event at $z \sim 1.7$. 

\begin{table}
  \begin{tabular}{  | c | c | }
    \hline
    Infall Mass & $f_{\rm debris}$ \\
    \hline
    \hline              
    $> 10^{10} \Msun$ & 0.12 \\
    $10^9 - 10^{10} \Msun$ & 0.42 \\
    $10^8 - 10^9 \Msun$ & 0.21 \\
    $10^7 - 10^8 \Msun$ & 0.16 \\
    $10^6 - 10^7 \Msun$ & 0.061 \\
    $< 10^6 \Msun$ & 0.027 \\
    \hline
  \end{tabular}
  \caption{The fraction of the debris flow contributed by halos in the given mass range.}
  \label{table: fdebris_by_mass}
\end{table}

To some degree, the larger fractional contribution of more massive individual halos is simply a result of their having more material to lose. In principle, it would be possible for the far greater number of low infall mass subhalos to contribute more to the debris flow in aggregate than the more massive ones. This turns out not to be the case: the majority ($>53$\%) of debris flow is contributed by subhalos with infall masses greater than $10^9 \Msun$, and halos with infall masses below $10^7 \Msun$ only contribute $<10$\% (see Table~\ref{table: fdebris_by_mass}). This result gives us confidence that our results are not highly sensitive to the numerical resolution of the VL2 simulation.

\section{Direct Detection Phenomenology}
\label{sec: model}

Direct detection experiments are sensitive to the scattering of dark matter particles off a target nucleus.  The recoil spectrum measured by these experiments depends on the distribution of dark matter speeds and is thus sensitive to the presence of local velocity substructure in the form of streams or debris flow.  The implications of tidal streams for direct detection experiments have been explored by~\cite{ Freese:2003tt, Gelmini:2004gm, Kuhlen:2009vh, Lang:2010cd, Alves:2010pt}, and more recently by~\cite{Natarajan:2011gz} in light of the CoGeNT anomaly~\cite{Aalseth:2010vx, Aalseth:2011wp}.  In this section, we will derive a semi-analytic model for the recoil spectrum of debris flows.  The phenomenological model presented here is a function of a single parameter and can easily be used to find the expected spectrum of events from scattering off the debris flow. 

For a direct detection experiment with target nucleus of mass $m_N$, the differential scattering rate per unit detector mass is~\cite{Lewin:1995rx}
\begin{equation}
\frac{dR}{dE_R} = \frac{\rho_0}{m_N m_{\text{dm}}} \sigma(E_R) g(v_{\text{min}}),
\label{eq: rate}
\end{equation}
where $\rho_0$ ($\approx 0.3$ GeV/cm$^3$) is the local dark matter density, $m_N$ is the nuclear mass, $E_R$ is the nuclear recoil energy, $\sigma(E_R)$ is the energy-dependent scattering cross section, and $g(v_{\text{min}})$ is a function of the detector's threshold speed. The differential scattering rate is sensitive to the distribution of dark matter speeds in the Earth frame $f(v)$, and thus depends on whether the dark matter is virialized or in a stream or debris flow.  The relevant quantity is 
\begin{equation}
g(v_{\text{min}}) = \int_{v_\text{min}} \frac{f(v)}{v} dv,
\label{eq: gvmin}
\end{equation}
where the threshold speed $v_{\text{min}}$ is given by $\sqrt{m_N E_R/2 \mu^2}$ for elastic scattering.  If the scattering is dominated by a Maxwellian distribution $f(v) \propto v^2 e^{-v^2/v_0^2}$ in the Galactic frame, the expected recoil spectrum is exponentially falling~\cite{Lewin:1995rx}.  If, in contrast, the local dark matter is dominated by a stream, then the scattering rate is constant up to a recoil energy corresponding to $|\vec{v}_{\rm stream} - \vec{v}_e|$~\cite{Freese:2003tt}, where $\vec{v}_e$ is given in Eq.~\ref{eq: vEarth}.

The particles in the debris flow have speeds characterized by the distribution function 
\begin{equation}
f(v)  = \frac{1}{N} \frac{dN}{dv} = \frac{1}{N} \frac{dN}{d\cos \theta_e} \frac{d\cos\theta_e}{dv} 
\end{equation}
in the Earth frame, where N is the total number of debris particles and $\theta_e$ is the angle between the velocities of the flow particles in the Galactic frame and the direction of Earth's motion.  This angle is related to the Earth-frame velocities through   
\begin{equation}
v^2 = v_{\text{flow}}^2 + v_e(t)^2 - 2 v_{\text{flow}} v_e(t) \cos \theta_e,
\label{eq: boost}
\end{equation}
where $v_{\text{flow}}$ is the speed of the debris flow in the Galactic frame.
\begin{figure}[tp] 
   \centering
\includegraphics[width=6.5in]{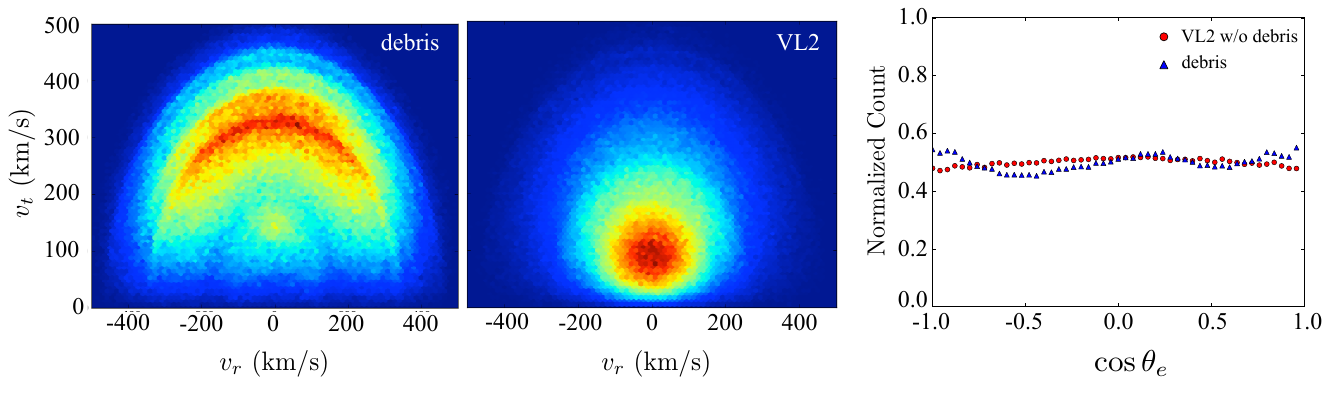}
   \caption{Tangential vs radial velocity components (km/s) of dark matter within a radial shell $7.5 < r < 9.5$ kpc in the Galactic frame.  On the left, the distribution for dark matter debris and in the middle, the distribution for all VL2 particles in this radial shell.  The right panel shows the distribution of debris particles (blue triangles) and all VL2 particles without the debris contribution (red circles) as a function of $\cos\theta_e$, where $\theta_e$ is the angle between the velocities of the particles in the Galactic frame and the direction of Earth's motion.}
   \label{fig: vrvtdistribution}
\end{figure}
A complete expression for f(v) depends on how the debris particles are distributed as a function of $\cos\theta_e$.  Figure~\ref{fig: vrvtdistribution} shows the tangential and radial Galactic-frame velocity distributions for the debris (left) and for all VL2 particles (middle) in a 7.5--9.5 kpc radial shall.  The right panel shows the distribution of debris particles as a function of $\cos\theta_e$.  The results show that  the debris flow is nearly uniformly distributed (isotropic) in $\cos\theta_e$, with $dN/d\cos\theta_e = N/2$.

To proceed, we make two simplifying assumptions. First, we neglect any dispersion in the Galacto-centric speed of the debris flow and treat its distribution as a delta function centered on $v_{\rm flow}$. Secondly, we assume that the debris flow is isotropic.  The true distribution of the debris flow's radial and tangential velocity components exhibits non-zero dispersion and a small tangential bias, but, as we show below, our simplified model nevertheless captures the main features of the recoil spectrum.

\begin{figure}[b] 
 \centering
  \includegraphics[width=6.5in]{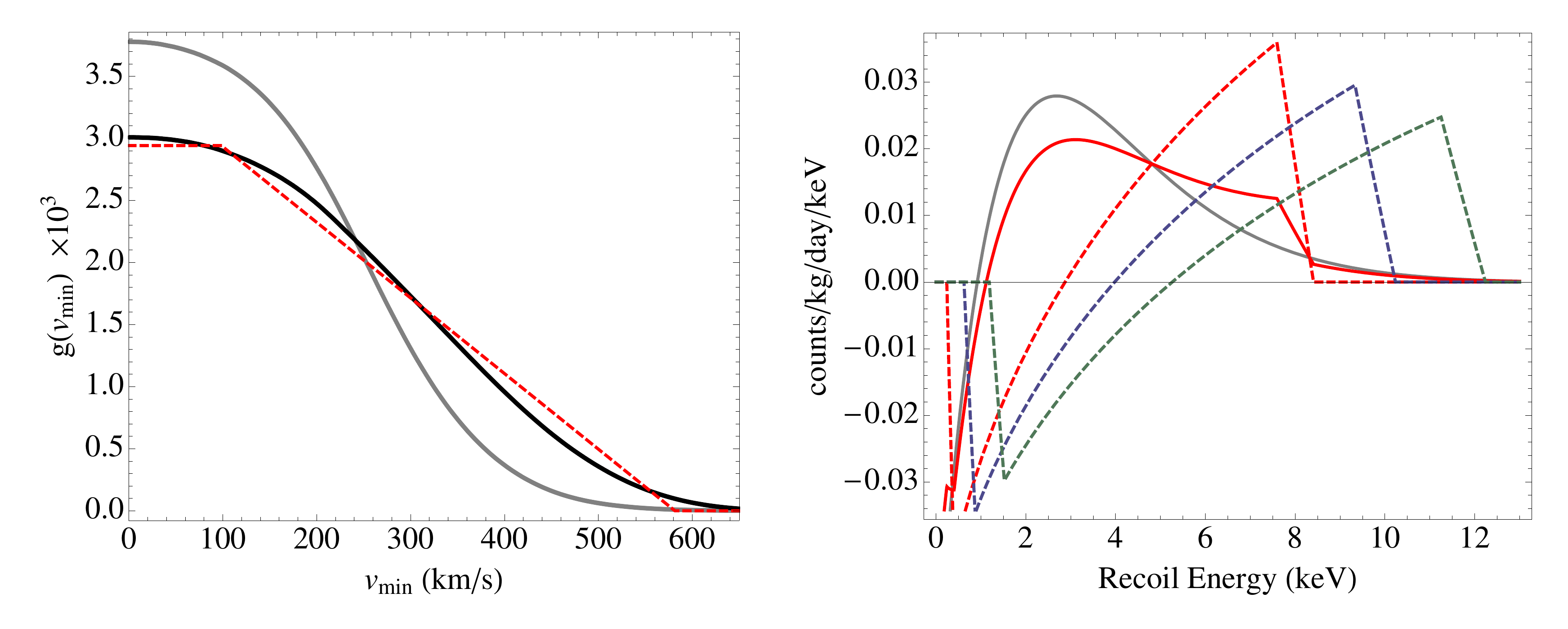}
  \caption{Left: $g(v_{\text{min}})$ for the debris flow (black) in a 7.5--9.5 kpc spherical shell in VL2.  The gray line represents the same distribution for all other particles in the same VL2 shell.  The dashed red line is the prediction of the semi-analytic model described in the text for $v_{\text{flow}} =  340$ km/s. These distributions are shown for $t_{\text{max}} = \text{ June 2}$. Right: The modulated amplitude for a 10 GeV dark matter elastically scattering off a Germanium target with cross section $10^{-41} \text{cm}^2$.  The gray line is the distribution for a Maxwellian distribution with $v_0 = 220$ and $v_{\text{esc}} = 550 \text{ km/s}$, while the dashed lines show the spectrum for debris flows with $v_{\text{flow}} = $ 340 (red), 400 (blue), and 460 (green) km/s.  The solid red line is the distribution for scattering off of both a Maxwellian and 340 km/s debris flow, with relative density given by VL2.  The modulated amplitude is half the difference in rate at peak (June) and minimum (December).}
  \label{fig: gvmin}
\end{figure}

With these assumptions, the Earth-frame speed distribution function of the debris flow is given by 
\begin{equation}
f_{\text{flow}}(v) = 
\begin{cases}
\frac{1}{2} \frac{v}{v_{\text{flow}} v_e(t)} & {\rm if} \; (v_{\rm flow} - v_e) < v < (v_{\rm flow} + v_e), \\
0 & {\rm otherwise}.
\end{cases}
\end{equation}
Substituting this into (\ref{eq: gvmin}) and integrating, we find that the recoil spectrum for the debris flow is proportional to
\begin{equation}
g(v_{\text{min}}) =
\begin{cases}
 \frac{1}{v_{\text{flow}}} & {\rm if} \; v_{\text{min}} < (v_{\text{flow}} -v_e), \\
 \frac{v_{\text{flow}} + v_e - v_{\text{min}}}{2 v_{\text{flow}} v_e} & {\rm if} \; (v_{\text{flow}} - v_e) < v_{\text{min}} < (v_{\text{flow}} + v_e), \\
 0 & {\rm if} \; v_{\text{min}} > (v_{\text{flow}} + v_e).
\end{cases}
\end{equation} 
The only input parameter in this equation is the speed of the debris flow in the Galactic frame.  The left panel of Figure~\ref{fig: gvmin}  shows the semi-analytic model for $g(v_{\text{min}})$ for $v_{\text{flow}} = 340$ km/s (dashed red).  Overlaid on the same plot is the distribution obtained directly from the VL2 simulation for the debris flow (solid black) and all other particles (solid gray) from 7.5--9.5 kpc.
The semi-analytic model captures the main features of the debris flow distribution remarkably well, even though it ignores the velocity dispersion and small tangential bias of the flow.

Because debris flow particles have relatively large speeds compared to the virialized component of the halo, they mainly contribute to nuclear recoils with large energies.  To illustrate this, let us consider the recoil energy spectrum for a 10 GeV elastically scattering dark matter particle.  The scattering rate for this low-mass dark matter candidate is given by Eq.~\ref{eq: rate}, with an energy-dependent cross section~\cite{Jungman:1995df}
\begin{equation}
\sigma(E_R) = \frac{m_N \sigma_N}{2 \mu^2} \frac{(f_p Z + f_n(A-Z))^2}{f_p^2} | F_H(E_R)|^2,
\end{equation}
where the detector target has charge $Z$ and atomic number $A$, $\mu$ is the reduced mass of the dark matter-nucleus system, $\sigma_N$ is the cross section for the dark matter-nucleus interaction at zero momentum transfer (10$^{-41} \text{ cm}^2$ for this example), and $f_{p,n}$ are the couplings to the proton and neutron, respectively.  We will take $f_p = f_n = 1$ for the rest of this section.  The Helm form factor $F_H(E_R)$ accounts for the loss of coherence at large momentum transfer~\cite{Helm:1956zz}.  

The right panel of Figure~\ref{fig: gvmin} shows the recoil energy spectrum of the modulated amplitude (half the difference between the maximum rate in June and the minimum in December) for this dark matter candidate scattering off a Germanium target.  The gray line is the spectrum assuming a Maxwellian distribution with $v_0 = 220 \text{ km/s}$ and $v_{\text{esc}} = 550 \text{ km/s}$, while the dashed lines are the distributions for a debris flow with Galacto-centric speeds of 340, 400, and 460 km/s (red, blue, green, respectively).  The total scattering rate will have contributions from both the virialized and unvirialized components of the halo.  Therefore, the total rate is a sum of the rates from  individual halo components - i.e., from the Maxwellian component, $R_{\text{MB}}$, and the debris flow component, $R_{\text{debris}}$:
\begin{equation}
\frac{d R_{\text{total}}}{dE_R} = (1 - \epsilon(0) ) \frac{dR_{\text {MB}} }{dE_R}+ \epsilon(0)  \frac{dR_{\text{debris}}}{dE_R}.
\end{equation}
The relative contribution from either component depends on the relative density fraction $\epsilon(0) = N_{\text{debris}}/N_{\text{tot}}$ between the debris flow and the total number of halo particles, which in the VL2 simulation is 0.22 (see Figure~\ref{fig: gvminplot}). The solid red line in the right panel of Figure~\ref{fig: gvmin} shows the recoil spectrum of the modulated amplitude when scattering occurs off of both a Maxwellian and 340 km/s debris flow, with $\epsilon(0)=0.22$.  Clearly, the presence of the debris flow leads to more significant modulation at recoil energies where a Maxwellian distribution would give little contribution.

It is clear, then, that the density and speed of the debris flow can have important implications for the expected distribution of events in direct detection experiments.  If the dark matter has a large scattering threshold, such as in light elastic dark matter or inelastic dark matter~\cite{TuckerSmith:2001hy}, it may be particularly sensitive to the presence of the debris flow.  Both of these scenarios have received attention recently, in light of conflicting results from current experiments.  The tightest limit for spin-independent scattering interactions is currently set by XENON100~\cite{Aprile:2011hi}, and improves upon bounds from CDMS~\cite{Ahmed:2009zw, Akerib:2005kh}, EDELWEISS~\cite{Edelweiss:2011cy}, XENON10~\cite{Angle:2007uj, Angle:2009xb}, CRESST~\cite{Angloher:2004tr}, and ZEPLIN~\cite{Alner:2007ja, Lebedenko:2008gb, Akimov:2011tj}.  Despite the null results from these experiments, the DAMA collaboration reports a $9\sigma$ annual modulation signal~\cite{Bernabei:2008yi, Bernabei:2010mq} and the CoGeNT experiment reports a 2.8$\sigma$ modulation~\cite{Aalseth:2010vx}.  The CRESST experiment has also claimed an excess of events that cannot be explained with background estimates~\cite{Angloher:2011uu}.  The three anomalies can be made consistent with each other if the dark matter is light $\OO(10\text{ GeV})$ and scatters off a non-Maxwellian distribution~\cite{Frandsen:2011gi}.  Debris flows may also be able to explain the observed modulation in CoGeNT at unexpectedly large energies~\cite{Fox:2011px}.  We caution, however, that a non-Maxwellian velocity distribution by itself does not appear to be sufficient to reconcile these signals with the exclusion limits from the CDMS low-threshold analyses~\cite{Ahmed:2010wy, Ahmed:2012vq, Collar:2011kf}, XENON10 S2 analysis~\cite{Angle:2011th}, and XENON100 results~\cite{Aprile:2011hi} (see \cite{Frandsen:2011gi}).  

\begin{figure}[htbp] 
   \centering
   \includegraphics[width=0.32\textwidth]{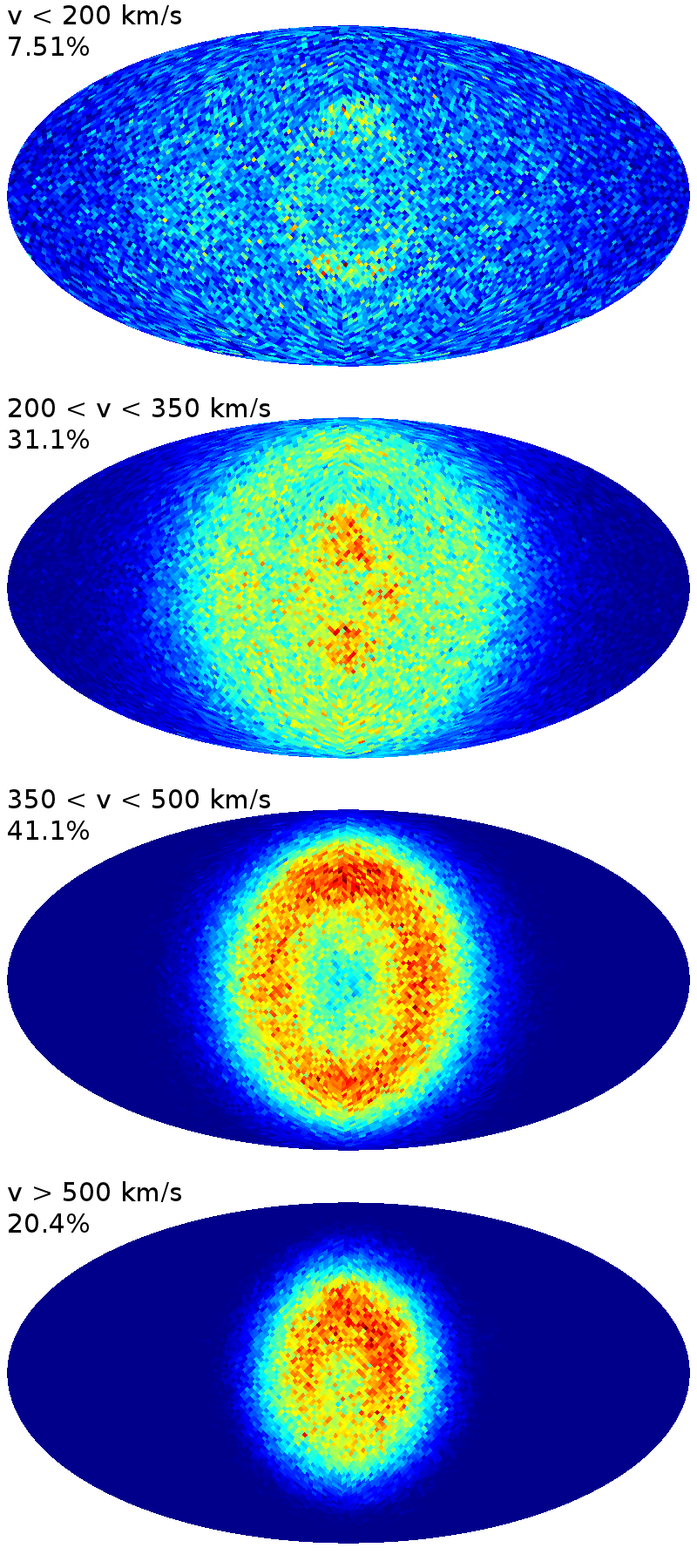}
   \includegraphics[width=0.32\textwidth]{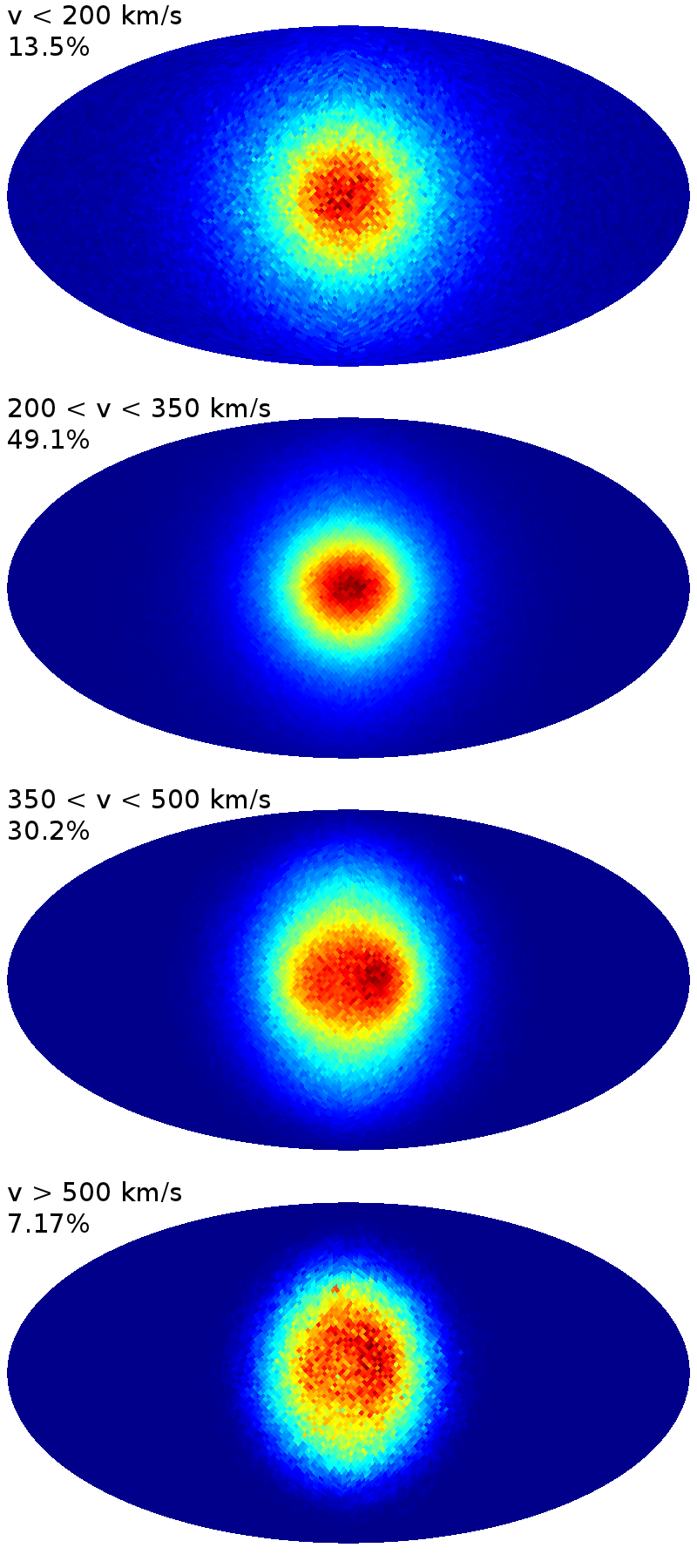}
   \includegraphics[width=0.32\textwidth]{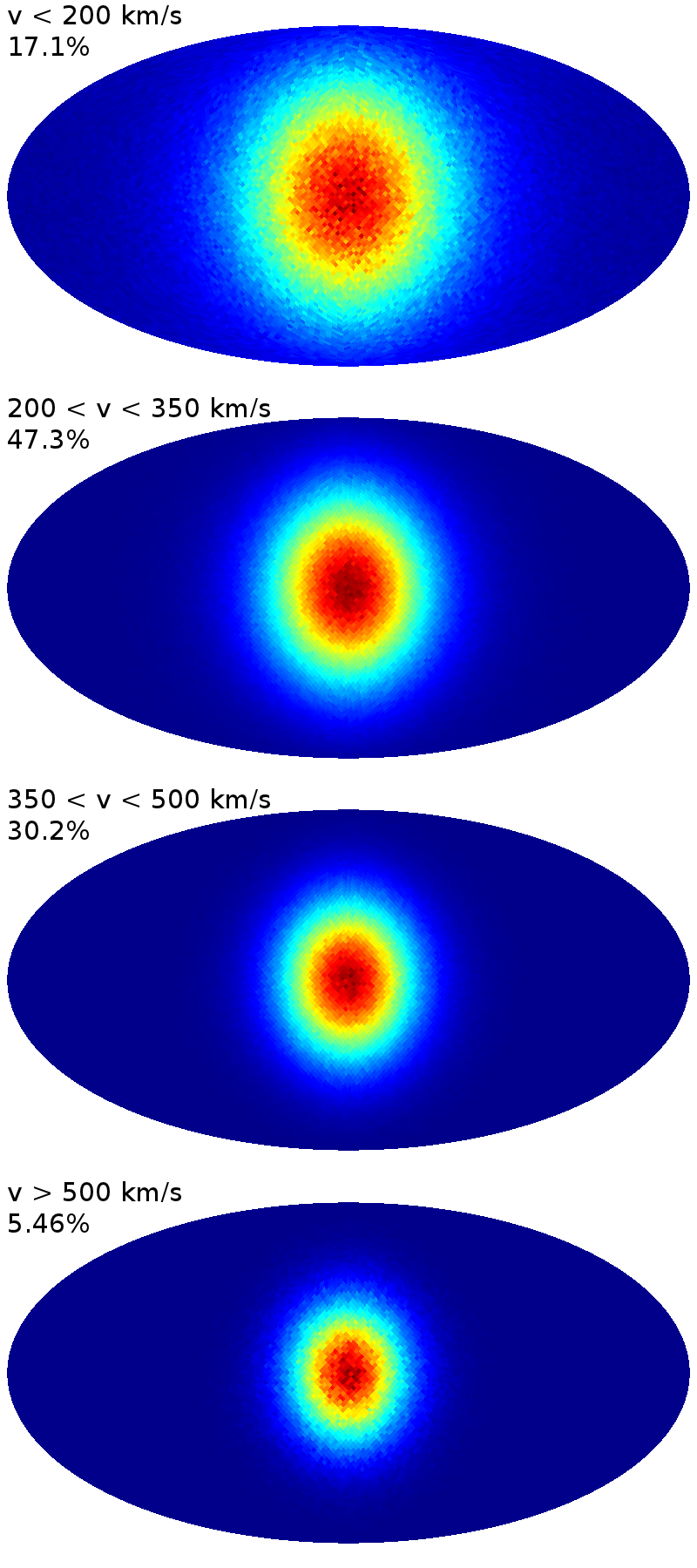}
   \caption{Mollweide projections of the distributions of incidence direction of debris particles (left), all particles (middle), and a $10^7$ particle realization of a Maxwellian halo (right). The coordinate system is chosen such that the Galactic disk normal is aligned with the simulation's $\hat{y}$-direction, and the direction anti-parallel to the Earth's motion is in the center of the projection. From top to bottom, the rows show the distributions for particles with Earth-frame speeds $< 200 \kms$, $200 - 350 \kms$, $350 - 500 \kms$, and $> 500 \kms$.}
   \label{fig:example}
\end{figure} 

The presence of velocity substructure could be even more important for experiments that are sensitive to the direction of the scattering dark matter particles \cite{Kuhlen:2009vh}, rather than just their speed. Indeed, directionally sensitive detectors, such as DRIFT \cite{Burgos:2007zz}, DMTPC \cite{Sciolla:2009fb}, MIMAC \cite{Santos:2011kf}, and NEWAGE \cite{Miuchi:2010hn} (see \cite{Ahlen:2009ev} for a summary of the current state of experimental efforts), require large recoil energies in order to follow the recoil tracks, which are typically only a few millimeters in length. These experiments are thus quite likely to be impacted by the non-Maxwellian velocity structure arising from debris flows.

To investigate the expected directional signatures of the debris flow in more detail, we present in Figure~\ref{fig:example} Mollweide projections of the distribution of incidence directions for debris particles (left), for all particles (center), and for a Maxwellian halo (right). The coordinate system is chosen such that the direction anti-parallel to the Earth's motion corresponds to the center of the maps. We show the distributions split into four distinct Earth-frame speed bins, in order to demonstrate trends with recoil energy. The incidence directions of the debris flow particles are distributed more broadly and less uniformly than for the Maxwellian halo, and they exhibit remarkable ring-like structures, most pronounced in the 350 km/s $< v <$ 500 km/s bin.\footnote{The detailed morphology and strength of these features depends somewhat on the orientation of the Galactic disk plane, which is not specified in the purely dark matter VL2 simulation.} Such features arise because the debris flow is peaked at one Galacto-centric speed ($\sim 350$ km/s), but is nearly isotropic in direction. Debris flow particles that happen to be traveling in the direction anti-parallel to Earth are boosted out of the 350--500 km/s bin, and far fewer lower speed particles are boosted into this bin. The result is a hole in the center of the distribution. Similar effects occur in the other speed bins.

Experimentally, there is no way to determine whether a given recoil event originated with a debris flow or a relaxed halo particle, and so it perhaps makes more sense to look at the combined distribution for all particles, as shown in the middle column of Figure~\ref{fig:example}. Now, the ring-like features are washed out by the dominating relaxed halo component, but in the two highest speed (i.e. recoil energy) bins a pronounced asymmetry is still visible. Comparing to the equivalent Maxwellian distributions (in the right column), it is apparent that the debris flow has two effects at high speeds: (i) it broadens the distribution of incidence directions and (ii) the peak of the distribution (the hotspot direction) can be shifted away from the direction anti-parallel to the Earth's motion. The latter effect is due to anisotropy in the direction of debris flow particles.

\section{Conclusions}
\label{sec: conclusion}

This work presents a detailed analysis of the properties of dark matter debris flow in the VL2 simulation.  Debris flow is an example of spatially-uniform velocity substructure that consists of overlapping sheets, streams, plumes and shells created by dark matter that is tidally stripped from subhalos falling into the Milky Way.  Subhalos that contribute dominantly to debris flow typically  have large infall mass ($\gtrsim 10^9$ M$_{\odot}$) and make numerous ($\gtrsim 10$) pericenter passages, with a minimum pericenter distance within 8 kpc.  

Debris flow is distinct from dark matter streams.  Although both arise from tidal disruption of satellites, streams are dynamically colder than debris flow and have not had time to spatially mix.  Streams consist of particles that are spatially confined and coherent in velocity space.  In contrast, debris flow is spatially-mixed over a large volume, yet retains distinctive velocity behavior because it is not completely virialized.  In VL2, the debris flow has a speed peaked at a magnitude of $\sim 340$ km/s.  

Debris flow is ubiquitous in the solar neighborhood; approximately 20\% of all dark matter particles in VL2 between 7.5--9.5 kpc are identified as debris flow.  This fraction increases to 50\% for particles with speeds greater than 450 km/s, and rises to 80\% at 600 km/s. The prevalence of debris flow makes it highly relevant for direct detection experiments.  In particular, if the dark matter has a large minimum scattering threshold, then direct detection experiments will be sensitive to its presence.  The recoil spectrum is different from that expected for a standard Maxwellian distribution, with more scattering events at large nuclear recoil energies.  Our simple parametrization for the debris flow recoil spectrum allows one to analytically determine the deviations from a Maxwellian expectation for a debris flow of given speed and density.    

Although the primary focus of this work has been to study debris flow in the context of dark matter, we conclude with some preliminary thoughts on the relevance of debris flow to the stellar halo.  The dense cores of subhalos were the site of star formation billions of years ago and these stars are tidally-stripped, along with dark matter, as the subhalos fall into the Milky Way.  As a result, the distribution of stars in the halo is not smooth, and exhibits phase-space features that are correlated with accretion events in the Galaxy~\cite{Johnston:1996sb, Johnston:1997fv, Bullock:2005pi, Johnston:2008fh, Harding:2000zt,Helmi:1999ks}.  A tidal stream is an example of such a feature, and evidence for streams has been found using deep photometric wide-field surveys, such as SDSS~\cite{Ivezic:2000ua, Yanny:2000ty}, the Spaghetti Survey~\cite{Morrison:2000gp} and the Two Micron All Sky Survey~\cite{2MASS} (see~\cite{Helmi:2008eq} for a review).  

The presence of stellar streams strongly suggests that debris flow should also be present and potentially detectable.  Ideally, a search for spatially-uniform velocity substructure requires complete kinematic information of stars.  The upcoming GAIA satellite~\cite{Perryman:2001sp} will obtain the largest and most accurate sample of proper motions in the solar neighborhood to date and will therefore be an integral step in mapping out the velocity domain.  In the meantime, a study using the position and radial velocity measurements of metal-poor main sequence turnoff stars in 137 SEGUE lines of sight has found evidence for velocity substructure~\cite{Schlaufman:2009jv}.  This study identified 10 high-confidence and 21 lower-confidence\footnote{They expect 3 false positives in this subset.} detections, referred to as ECHOS (Elements of Cold HalO Substructure), within 17.5 kpc of the sun.  Each detection consists of $\OO(20)$ stars uniformly distributed along a large patch of sky with a radial velocity distribution that differs from the expected background.  In addition, the ECHOS are chemically distinct from the kinematically smooth stellar halo background, strongly suggesting a separate origin~\cite{Schlaufman:2011kf}.  Because the detections are spread out over large areas of the sky, they do not exhibit a stream-like morphology.  Indeed, the morphology more closely resembles that of debris flow, and it will be useful to explore whether the ECHOS can be explained as the tidal debris of many infalling satellites.

Discovery of debris flow in the stellar halo would provide critical information for dark matter searches, suggesting that there are more high speed particles (with speeds in excess of the most probable speed) in the solar neighborhood than expected for a Maxwellian standard halo model. This would alter the expectation for the nuclear recoil spectrum and modulation fraction in direct detection experiments, as well as the angular distribution of events in directional experiments.  In light of a detection, the results from both the dark matter and stellar searches will shed light on the matter distribution in our Galaxy and its tumultuous merger history.

\section*{Acknowledgements}
We thank Jesse Thaler, David Weinberg, and Neil Weiner for useful discussions, and Jim Cline and Wei Xue for catching an error in our debris flow toy model.  ML acknowledges support from the Simons Postdoctoral Fellows Program.  This work was supported in part by the U.S. National Science Foundation, grant NSF-PHY-0705682, the LHC Theory Initiative, Jonathan Bagger, PI, and NSF grants OIA-1124453 (PI P.~Madau) and OIA-1124403 (PI A.~Szalay).

\bibliography{dirdetBibFile}

\end{document}